\documentclass[prl,twocolumn,floatfix,superscriptaddress]{revtex4-1}
\usepackage{dcolumn,amsmath}
\usepackage{graphicx}
\usepackage{bm}
\usepackage{hyperref}

\newcommand{\Eeff}{\ensuremath{E_{\rm eff}}}
\newcommand{\eEDM}{{\em e}EDM}

\newcommand{\Eref}[1]{Eq.~(\ref{#1})}

\begin{document}
\title{Enhanced effect of CP-violating nuclear magnetic quadrupole moment in HfF$^+$ molecule}

\author{L.V.\ Skripnikov}\email{leonidos239@gmail.com}
\author{A.V.\ Titov}
\homepage{http://www.qchem.pnpi.spb.ru}

\affiliation{National Research Centre ``Kurchatov Institute'' B.P. Konstantinov Petersburg Nuclear Physics Institute, Gatchina, Leningrad District 188300, Russia}
\affiliation{Saint Petersburg State University, 7/9 Universitetskaya nab., St. Petersburg, 199034 Russia}

\author{V. V. Flambaum}
\affiliation{School of Physics, The University of New South Wales, Sydney
NSW 2052, Australia}

\date{12.01.2017}

\begin{abstract}
HfF$^+$ cation is a very promising system to search for the electron electric dipole moment (EDM), and  corresponding experiment is carried out by E. Cornell group. Here we theoretically investigate the cation to search for another T,P-odd effect -- the nuclear magnetic quadrpole moment (MQM) interaction with electrons. We report the first accurate  {\it ab~initio} relativistic electronic structure calculations of the molecular parameter $W_M$=0.494 $\frac{10^{33}\mathrm{Hz}}{e~{\rm cm}^2}$ that is required to interpret the experimental data in terms of the MQM of Hf nucleus. For this we have implemented and applied the combined Dirac-Coulomb(-Gaunt) and relativistic effective core potential approaches to treat electron correlation effects from all of the electrons and to take into account high-order correlation effects using the coupled cluster method with single, double, triple and noniterative quadruple cluster amplitudes, CCSDT(Q). We discuss interpretation of the MQM effect in terms of the strength constants of T,P-odd nuclear forces, proton and neutron EDM, QCD parameter $\theta$ and quark chromo-EDM.
\end{abstract}

\maketitle

\section{Introduction}

HfF$^+$ cation is a very promising system to search for the electron electric dipole moment (\eEDM) \cite{Cossel:12, Cornell:13, Petrov:07a, Petrov:09b, Fleig:13, Meyer:06a, Skripnikov:08a, Le:13} (see also \cite{Meyer:08,ACME:14a,FDK14,Sandars:64, Sandars:67, Gorshkov:79, Sushkov:78}).
At present E. Cornell's group prepares the ion trap experiment on the cation \cite{Meyer:06a,Meyer:08}.
In contrast to the $^{232}$ThO molecule which was used to obtain the best current limit on \eEDM\ \cite{ACME:14a} one can  use available stable isotope of Hf, e.g. $^{177}$Hf
to search for the magnetic quadrupole moment of the $^{177}$Hf nucleus in the $^{177}$HfF$^+$ cation \cite{FDK14}. This is because the $^{177}$Hf nucleus posses nuclear spin  I$>1/2$ \cite{FKS84b,KhMQM,F94} while $^{232}$Th has $I=0$.

As was shown in \cite{F94} MQM can be strongly enhanced due to the collective nuclear effect. Below we study this effect for the case of Hf nucleus.

Electronic structure of HfF$^+$ cation has been previously studied in Refs.~\cite{Petrov:07a,Petrov:09b,Cossel:12,Fleig:13} for the \eEDM\ problems -- calculation of the effective electric field (\Eeff) which is required to interpret the experimental energy shift in terms of \eEDM.
In Refs.~\cite{Petrov:07a,Petrov:09b} the two-step relativistic effective core potential approach was used.
In Ref.~\cite{Fleig:13} a direct approach within the Dirac-Coulomb Hamiltonian was applied.
In the present paper we follow the new combined Dirac-Coulomb(-Gaunt) and two-step relativistic pseudopotential scheme \cite{Skripnikov:16b} to study the electronic part of the problem of calculation of the interaction between the MQM of Hf nucleus and electrons of HfF$^+$ in the first excited $^3\Delta_1$ state of HfF$^+$ cation. This scheme allows one to treat all of the important effects including correlation of the inner-core electrons.

\section{Theory}

Qualitatively the effect under consideration corresponds to the interaction of the nuclear magnetic quadrupole moment with  the gradient of the magnetic field produced by electrons.
This is the T,P-odd interaction which mixes states of opposite parity in atoms and molecules \cite{FKS84b,KhMQM}.
Relativistic Hamiltonian of the interaction is given by the following expression \cite{FKS84b,Kozlov:87,GFreview}:
 \begin{align}\label{hamq}
 H^{\rm MQM}  &=
 -\frac{  M}{2I(2I-1)}  T_{ik}\frac{3}{2} \frac{[\bm{\alpha}\times\bm{r}]_i r_k}{r^5},
 \end{align}
where Einstein's summation convention is implied, 
$ \bm{\alpha}=
  \left(\begin{array}{cc}
  0 & \bm{\sigma} \\
  \bm{\sigma} & 0 \\
  \end{array}\right)
$
are the 4x4 Dirac matrices,
 $\bm{r}$ is the displacement of the electron from the Hf nucleus, $\bm I$ is the nuclear spin, $M$ is the nuclear MQM,
\begin{align}\label{eqaux1}
M_{i,k}=\frac{3M}{2I(2I-1)}T_{i,k}\, \\
 T_{i,k}=I_i I_k + I_k I_i -\tfrac23 \delta_{i,k} I(I+1)\,.
 \end{align}

In the subspace of $\pm \Omega$ states ($\Omega= \langle\Psi|\bm{J}\cdot\bm{n}|\Psi\rangle$,
$\bm{J}$ is the total electronic momentum, $\Psi$ is the {\it electronic} wave function for the considered $^3\Delta_1$ state of HfF$^+$) expression (\ref{hamq}) is reduced to the following effective molecular Hamiltonian~\cite{FKS84b}:
 \begin{align}\label{eq0}
H^{\rm MQM}_\mathrm{eff} &=
 -\frac{W_M  M}{2I(2I-1)} \bm S^{\prime} \hat{\bm T} \mathbf{n}
 \,,
 \end{align}
where $\mathbf{n}$ is the unit vector along the molecular axis $\zeta$ directed from Hf to F, $\bm S^{\prime}$ is the effective electron spin~\cite{Kozlov:95}
defined by the following equations: 
$\mathbf{S}^{\prime}_\mathbf{\zeta}|\Omega> = \Omega|\Omega>$,
$\mathbf{S}^{\prime}_{\pm}|\Omega=\pm 1> = 0$ \cite{Kozlov:87, Dmitriev:92}, $S{=}|\Omega|{=}1$.
$W_M$ parameter is defined by the following equation:
\begin{align}
  \label{WM}
W_M= 
\frac{3}{2\Omega} 
   \langle
   \Psi\vert\sum_i\left(\frac{\bm{\alpha}_i\times
\bm{r}_i}{r_i^5}\right)
 _\mathbf{\zeta} r_\mathbf{\zeta} \vert
   \Psi\rangle\ .
 \end{align}

As was shown in Ref.~\cite{Skripnikov:14a} for a completely polarized molecule the energy shift due to MQM interaction is:
\begin{eqnarray}
\label{shiftM}
\delta_M(J,F) = (-1)^{I+F}C(J,F) M W_M \Omega\ , \\
C(J,F)= \frac{(2J+1)}{2}
\frac{
    \left(
    \begin{array}{ccc}
    J &  2 &  J \\
   -\Omega & 0 & \Omega
    \end{array}
    \right)
    }
    {
    \left(
    \begin{array}{ccc}
    I &  2 &  I \\
   -I & 0 & I
    \end{array}
    \right)
    }
    \left\{
    \begin{array}{ccc}
    J &  I &  F \\
    I &  J & 2
    \end{array}
    \right\},
\end{eqnarray}
where $(...)$ means elements with 3j$-$symbols and $\{...\}$ are those with 6j$-$symbols \cite{LL77}, $F$ is the total angular momentum and $J$ is the number of rotational level.
Note, that $\delta_M$ depends on $J$ and $F$ quantum numbers. Besides $H^{\rm MQM}_\mathrm{eff}$ has non-zero off-diagonal matrix elements on $J$ quantum number (between different rotational levels).
This should be taken into account when mixing of different rotational levels become significant. In \Eref{shiftM} this effect is neglected.
For $^{177}$HfF$^+$ ($I{=}7/2$) and ground rotational level {$J{=}1$} \Eref{shiftM} gives the MQM energy shifts, $\left| \delta(J,F) \right|$, equal to $0.107 W_M M, \ 0.143 W_M M, \ 0.05 W_M M$ for $F= 5/2,7/2,9/2$, correspondingly.

\section{Calculation of the Nuclear Magnetic Quadrupole moment}
The angular momentum  $I$ of a spherical nucleus is determined by a valence nucleon. In the single-valence-nucleon model the nuclear MQM is given by the following expression:
\begin{align}\label{M}
M=[d-2\cdot 10^{-21} \eta(\mu-q)  (e \cdot {\rm cm})] \lambda_p (2I-1)t_I,
 \end{align}
where $t_I=1$ for $I=l+1/2$ and $t_I=-I/(I+1)$ for $I=l-1/2$, $I$ and $l$ are the total and orbital angular momenta of a valence nucleon,
 $\eta$ is the dimensionless strength constant of the T,P-odd nuclear potential $\eta G/(2^{3/2} m_p) (\sigma \cdot \nabla \rho)$ acting on the valence nucleon,  $\rho$ is the total nucleon number density,  the nucleon magnetic moments are $\mu_p=2.79$ for valence proton  and $\mu_n=-1.91$ for valence neutron, $q_p=1$ and $q_n=0$, $\lambda_p=\hbar /m_pc=2.1 \cdot 10^{-14}$ cm.  The  contribution of the valence nucleon EDM $d$ was calculated in Ref. \cite{KhMQM} , the contribution of the T,P-odd nuclear forces was calculated
 in  \cite{FKS84b}. 
 Using a natural assumption that in any model of CP-violation the $\pi$ meson  exchange gives  significant contributions  it was  concluded in  \cite{FKS84b} that the  contribution of the T,P-odd nuclear forces to any T,P-odd nuclear moment  is 1-2 orders of magnitude larger than the contribution of the nucleon EDM \cite{FKS84b}. 

In a deformed nucleus the MQM  in the ``frozen'' frame (rotating together with a nucleus) may be estimated using the following formula \cite{F94}:
 \begin{align}\label{Mzz}
M^{\rm nucl}_{zz}=\sum M^{\rm single}_{zz} (I,I_z,l) n(I,I_z,l),
 \end{align}
where the sum goes over occupied orbitals,  $M^{\rm single}_{zz}(I,I_z,l)$ is given by Eqs.~(\ref{M}) and (\ref{eqaux1}), $T_{zz}=2 I_z^2 -\tfrac23 I(I+1)$, $n(I,I_z,l)$ are the orbital occupation numbers, which may be found in Ref. \cite{Bohr}. The sum over a complete shell gives zero; therefore, for shells more than half-filled, it is convenient to use hole numbers in place of particle numbers, using the relation $M^\mathrm{single}_{zz} (\mathrm{hole})=- M^\mathrm{single}_{zz}(\mathrm{particle})$. 

The nucleus $^{177}$Hf  has  the following occupation numbers: 13 neutron holes in the orbitals  $[\bar{l}_I, I_z] = [\bar{f}_{7/2}, -7/2]$,
 $[\bar{i}_{13/2},\pm 13/2, \pm 11/2, \pm 9/2]$, $[\bar{h}_{9/2}, \pm 9/2, \pm 7/2 ]$,  $[\bar{p}_{3/2}, \pm 3/2]$, and 8 proton holes $[\bar{d}_{3/2}, \pm 3/2 ]$, $[\bar{d}_{5/2}, \pm 5/2 ]$,  
 $[\bar{h}_{11/2}, \pm 11/2, \pm 9/2]$.

The MQM in the laboratory frame, $M\equiv M_{\rm lab}$, can be expressed via MQM in the rotating frame (\ref{Mzz}):
\begin{align}\label{Mlab}
\nonumber 
 M^{\rm lab}=\frac{I(2I-1)}{(I+1)(2I+3)} M^{\rm nucl}_{zz}=\\
\nonumber
(1.5 \eta_p- 1.1  \eta_n) \cdot 10^{-33}  (e \cdot {\rm cm}^2)\\
-(4.0 d_p +2.9 d_n) \cdot 10^{-13}  {\rm cm},
 \end{align}
where   $I=7/2$ 
is the  nuclear spin of $^{177}$Hf.

The T,P-odd nuclear forces are dominated by the $\pi_0$ meson exchange \cite{FKS84b}. Therefore, we may express the strength constants via strong $\pi NN$ coupling constant $g=13.6$ and T,P-odd $\pi N N$ coupling constants corresponding to the isospin channels  $T=0,1,2$:  $\eta_n= - \eta_p = 5\cdot 10^6 g (  {\bar g_1}{+}0.4{\bar g_2}{-}0.2{\bar g_0}) $ (see detailes in \cite{Skripnikov:14a}). 
As a result, we obtain 
\begin{align}\label{Mg}
M(g)= - [g (  {\bar g_1}+ 0.4 {\bar g_2}-0.2 {\bar g_0})  \cdot  1.0 \cdot 10^{-26} e \cdot {\rm cm}^2 .
 \end{align}
Possible CP-violation in the strong interaction sector is described by the  CP violation parameter ${\tilde \theta}$. According to Ref.~\cite{theta}  $g {\bar g_0}=-0.37 {\tilde \theta} $. This gives the following value of MQM for $^{177}$Hf:
\begin{align}\label{Mtheta}
M(\theta) = -7 \cdot 10^{-28} {\tilde \theta} \cdot e \cdot {\rm cm}^2 .
 \end{align} 


Finally, we can express MQM in terms of the quark chromo-EDM ${\tilde d_u}$ and  ${\tilde d_d}$ using the relations 
 $g {\bar g_1}=4.{\cdot}10^{15}( {\tilde d_u} - {\tilde d_d})/{\rm cm} $, $g {\bar g_0}=0.8 \cdot 10^{15}( {\tilde d_u} + {\tilde d_n})/{\rm cm} $ 
 \cite{PospelovRitzreview}:
\begin{align}\label{Md}
 M( {\tilde d}) = -4 \cdot 10^{-11} ( {\tilde d_u} - {\tilde d_d}) \cdot e \cdot {\rm cm} .
 \end{align}
The contributions of $d_p$  and  $d_n$ to MQM  in Eqs.~(\ref{Mg} -\ref{Md}) are from one to two orders of magnitude smaller than the contributions of the nucleon T,P-odd interactions.

\section{Electronic structure calculation details}
It follows from Eq.~(\ref{WM}) that $W_M$ parameter is mainly determined by the behavior of the electronic wave function in the region close to the heavy atom nucleus.
We call such parameters as the Atoms-In-Compounds characteristics or properties \cite{Skripnikov:15b,Titov:14a,Zaitsevskii:16a}. Other examples are the hyperfine structure interaction constants, effective electric field, chemical shifts, etc.
To compute such parameters we have previously developed the two-step method \cite{Titov:06amin,Skripnikov:15b,Skripnikov:16a} which allows us to avoid direct 4-component relativistic treatment.
In the first stage, one considers the valence (and outer-core) part of the molecular wave function within the generalized relativistic effective core potential (GRECP) method \cite{Titov:99,Mosyagin:10a,Mosyagin:16}. The inner-core electrons are excluded from the explicit treatment. The feature of this stage is that the valence wave functions (spinors) are smoothed in the spatial inner core region of a considered heavy atom. This leads to considerable computational savings. Some technical advantage is that one can also use very compact contracted basis sets \cite{Skripnikov:16b,Skripnikov:13a}. This is of crucial importance to treat high-order correlation effects.
Besides, one can exclude the spin-orbit term of the GRECP operator and consider scalar-relativistic approximation with a good nonrelativistic symmetry. Due to the corresponding savings one can use very large basis sets to consider basis set corrections and analyze its saturation.
At the second step, one uses the nonvariational procedure developed in \cite{Titov:06amin,Skripnikov:15b,Skripnikov:16a,Skripnikov:11a} to restore the correct 4-component  behavior of the valence wave function in the spatial core region of a heavy atom.
The procedure is based on a proportionality of the valence and low-lying virtual spinors in the inner-core regions of heavy atoms. The procedure has been recently extended to consider not only the atomic and molecular systems but also three-dimensional periodic structures (crystals) in Ref.~\cite{Skripnikov:15a}. GRECP and the restoration procedure were also successfully used for precise investigation of different diatomics~\cite{Lee:13a,Skripnikov:15b,Skripnikov:14c,Petrov:13,Kudashov:13,Kudashov:14,Skripnikov:09,
Skripnikov:15d,Petrov:14,Skripnikov:13c,Skripnikov:09a,Skripnikov:08a}.
The two-step method allows one to consider high-order correlation effects and large basis sets with rather modest requirements to computer resources in comparison to 4-component approaches. However, some uncertainty remains due to the impossibility to consider the full version of the GRECP operator in the currently available codes and neglect of the inner-core correlation effects. In Refs.~\cite{Skripnikov:16b,Skripnikov:17a} we suggested to combine the two-step approach and the direct relativistic Dirac-Coulomb(-Gaunt) approach to take advantages of both approaches.

Computational scheme of the molecular $W_M$ parameter (\ref{WM}) assumes evaluation of the following contributions:
(I) the main correlation contributions within the 52-electron 4-component Dirac-Coulomb coupled cluster with single, double and noniterative triple cluster amplitudes, CCSD(T), theory;
(II) the inner-core correlation contributions;
(III) correction on inclusion of the Gaunt interaction;
(IV) contribution of high-order correlation effects up to the coupled cluster with single, double, triple and noniterative quadruple amplitudes for the valence electrons within the 2-component (with spin-orbit effects included) two-step approach;
(V) calculation of the basis set correction for 52 outer electrons of HfF$^+$ within the scalar-relativistic two-step approach.

For step (I) we used the CVQZ basis set for Hf \cite{Dyall:07,Dyall:12} and aug-ccpVQZ basis set \cite{Dunning:89,Kendall:92} with two removed g-type basis functions for F. The inner-core electrons ($1s..3d$ of Hf) were excluded from the correlation treatment.  For the outer-core/valence correlation calculation we set cutoff equal to 50 Hartree for the virtual spinors.
The inner-core correlation contribution was calculated at the CCSD level as the difference between the $W_M$ values calculated with correlation of all 80 electrons of HfF$^+$ included into correlation treatment and with 52 electrons as in stage (I).
For these calculations we used the CVDZ \cite{Dyall:07,Dyall:12} basis set on Th and the cc-pVDZ~\cite{Dunning:89,Kendall:92} basis set on F. We set cutoff equal to 7000 Hartree for virtual molecular spinors in these calculations to be sure that the necessary correlation functions present in the one-electron spinor basis.
Correction at the step (III) has been calculated at the Hartree-Fock level.
In the stage (IV) 20 electrons of HfF$^+$ were correlated.
Correction was estimated as the difference in the calculated values of $W_M$ within the CCSDT(Q) versus the CCSD(T) method. For Hf we use slightly reduced version [12,16,16,10,8]/(6,5,5,1,1) of the basis set which was used in Refs.~\cite{Petrov:07a,Petrov:09b,Skripnikov:08a}. For F the ANO-I basis set \cite{Roos:05} reduced to [14,9,4,3]/(4,3,1) was used.
In the stage (V) we considered the influence of additional 7 $g-$, 6 $h-$ and 5 $i-$ basis functions on Hf (with respect to the basis functions of these types included in the CVQZ basis set, used at step (I)).
For stages (IV) and (V) we used the semilocal versions of 12-electron and 44-electron GRECP operators \cite{Petrov:07a,Petrov:09b,Skripnikov:08a,Mosyagin:10a,Mosyagin:16}.

In all the calculations the Hf$-$F internuclear distance in the $^3\Delta_1$ state was set to 3.41 Bohr \cite{Cossel:12}.

For the Hartree-Fock calculations and integral transformations we used the {\sc dirac12} code \cite{DIRAC12}. Relativistic correlation calculations were performed within the {\sc mrcc} code  \cite{MRCC2013}. For scalar-relativistic calculations we used the {\sc cfour} code \cite{CFOUR,Gauss:91,Gauss:93,Stanton:97}. 
The code to compute matrix elements of the MQM Hamiltonian has been developed in the present paper.

\section{Results and discussions}

The final value of $W_M$ is 0.494 $\frac{10^{33}\mathrm{Hz}}{e~{\rm cm}^2}$.

Inner-core contribution to the final value of $W_M$ is about 3\%. Gaunt contribution is about -1.6\%. High-order correlation effects give -0.3\%. This means that the convergence with respect to correlation effects is achieved.
Basis set correction on high-order harmonics is negligible (in contrast to the ThO case \cite{Skripnikov:16b}).
We estimate the accuracy of the final value of $W_M$ to be better than 4\%. The main uncertainly is due to omitting the ``interference'' of the Gaunt interaction and correlation effects.
Interestingly, the estimate of \cite{FDK14} appears to be rather close to our value though for the other considered systems the uncertainly was rather large, e.g. for ThF$^+$ the estimate from Ref.~\cite{FDK14} differs from that of ab-initio correlation calculation \cite{Skripnikov:15b} by about 3 times 
\cite{Note1}

The obtained $W_M$ in HfF$^+$ is very close to the value of $W_M$ in the ThF$^+$ \cite{Skripnikov:15b} and slightly smaller than  that in TaN \cite{Skripnikov:15c,Fleig:16a}. 
Note, however, that the HfF$^+$ cation is already under active investigation for the other T,P-odd effects and similar experimental technique can be used to search for the nuclear MQM.

One can express the MQM energy shift, $(-1)^{I+F}C(J,F) M W_M \Omega$ in terms of the fundamental CP-violating physical quantities $d_p$, $d_n$, $\tilde{\theta}$ and $\tilde{d}_{u,d}$ using Eqs.~(\ref{Mlab},\ref{Mtheta},\ref{Md}). For the lowest rotational level, for which the coefficient $|C(J{=}1,F{=}7/2)|=0.143$ 
reaches a maximum value, we have
\begin{align} 
\label{EprotonEDM} 
0.143 W_M  M  &=  
- \frac{10^{25}(2.8  d_p+2.0 d_n)}{e \cdot \mathrm{cm}} \cdot \mu{\rm Hz} 
\end{align} 
\begin{align}
\label{EshiftTheta}
0.143 W_M  M  &= -5.0 \cdot 10^{10} {\tilde \theta} \cdot \mu{\rm Hz} 
\end{align}   
\begin{align} 
\label{EshiftD} 
0.143 W_M  M  &= -2.8 \cdot \frac{10^{27}({\tilde d_u}-{\tilde d_d})}{\mathrm{cm}} \cdot \mu{\rm Hz}
\end{align} 
The current limits on $d_p$, $|{\tilde \theta}|$ and $|{\tilde d_u}{-}{\tilde d_d}|$ ($|d_p|~<~2.0~\cdot~10^{-25} e \cdot $cm, $|{\tilde \theta}| < 1.5 \cdot 10^{-10}$, $|{\tilde d_u}{-}{\tilde d_d}|<5.7 \cdot 10^{-27} $~cm ~\cite{Hg2016} correspond to the shifts $|0.143\ W_M M | < 6~\mu$Hz, $ 7~\mu$Hz and $ 16 ~\mu$Hz, respectively.

\section*{Acknowledgement}
Molecular calculations were partly performed on the Supercomputer ``Lomonosov''.
The development of the code for the computation of the matrix elements of the considered operators as well as  the performance of all-electron calculations were funded by RFBR, according to the research project No.~16-32-60013 mol\_a\_dk. Two-step GRECP calculations were performed with the support of President of the Russian Federation Grant No.~MK-7631.2016.2 and Dmitry Zimin ``Dynasty'' Foundation. V.F.  acknowledges support from the Australian Research Council and the Gutenberg Fellowship.


\end{document}